\documentclass[aps,preprintnumbers,amsmath,amssymb,nofootinbib]{revtex4}  
\usepackage{graphicx}
\usepackage{bm}
\usepackage{epsfig}
\usepackage{color}
\usepackage{float}

\begin{document}
\title{Revisiting nonfactorizable charm-loop effects in exclusive FCNC $B$-decays}
\author{
Anastasiia Kozachuk$^{a}$ and Dmitri Melikhov$^{a,b,c}$}
\affiliation{
$^a$D.~V.~Skobeltsyn Institute of Nuclear Physics, M.~V.~Lomonosov Moscow State University, 119991, Moscow, Russia\\
$^b$Institute for High Energy Physics, Austrian Academy of Sciences, Nikolsdorfergasse 18, A-1050 Vienna, Austria\\
$^c$Faculty of Physics, University of Vienna, Boltzmanngasse 5, A-1090 Vienna, Austria}
\date{\today}
\begin{abstract}
We revisit the calculation of nonfactorizable corrections induced by charm-quark loops in exclusive FCNC $B$-decays. 
For the sake of clarity, we make use of a field theory with scalar particles: this allows us to focus on the conceptual issues  
and to avoid technical complications related to particle spins in QCD. We perform a straightforward calculation 
of the appropriate correlation function and show that it requires the knowledge of the full generic three-particle 
distribution amplitude with non-aligned arguments, $\langle 0|\bar s(y)G_{\mu\nu}(x)b(0)|B(p)\rangle$. Moreover, 
the dependence of this quantity on the variable $(x-y)^2$ is essential for a proper account of 
the $\left(\Lambda_{\rm QCD}m_b/m_c^2\right)^n$ terms in the amplitudes of FCNC $B$-decays. 
\end{abstract}
\maketitle
\section{Introduction}
\label{Sec_introduction}
The interest in the contribution of virtual charm loops in rare FCNC semileptonic and radiative leptonic decays of the 
$B$-mesons is two-fold: (i) although CKM-suppressed, the effect of the virtual charm-quark loops, including the narrow charmonia 
states which appear in the physical region of the $B$-decay, has a strong impact on the $B$-decay observables \cite{neubert} 
thus providing an 
unpleasant ``noise'' for the analysis of possible new physics effects; (ii) it is known that in the charmonia region, 
nonfactorizable gluon exchanges dominate the amplitudes posing a challenging QCD problem. 

A number of theoretical analyses of nonfactorizable effects induced by charm-quark contributions has been published in 
the literature. We will mention here only those that are directly related to the discussion of this letter: 
In \cite{voloshin}, an effective gluon-photon local operator describing the charm-quark loop has been 
calculated for the real photon as an expansion in inverse charm-quark mass $m_c$ and applied to inclusive $B\to X_s\gamma$ decays; 
Ref.~\cite{buchalla} obtained a nonlocal effective gluon-photon operator for the virtual photon (i.e. without expanding in inverse 
powers of $m_c$) and applied it to inclusive $B\to X_s l^+l^-$ decays. 
In \cite{khod1997} nonfactorizable corrections in exclusive FCNC $B\to K^*\gamma$ decays using local OPE have been studied; 
in \cite{zwicky1,zwicky3}, these corrections have been analyzed with light-cone sum rules using local OPE for the photon-gluon 
operator and three-particle light-cone distribution amplitudes of $K^*$-meson.

As emphasized in \cite{voloshin,buchalla,ligeti,paz,hidr}, local OPE for the charm-quark loop leads to a power 
series in $\Lambda_{\rm QCD} m_b/m_c^2$. This parameter is of order unity for the physical masses of $c$- and $b$-quarks and 
thus corrections of this type require resummation. The authors of \cite{hidr} derived a different form of the nonlocal 
photon-gluon operator compared to \cite{buchalla} and evaluated its effect at small values of $q^2$ 
($q$ momentum of the lepton pair) making use of light-cone 3-particle DA (3DA) of the $B$-meson with the aligned arguments,
$\langle 0|\bar s(y)G_{\mu\nu}(uy) b(0)|B_s(p)\rangle$. 

The goal of this letter is to emphasize that the full consistent resummation 
of $\left(\Lambda_{\rm QCD} m_b/m_c^2\right)^n$ terms in the nonfactorizable amplitude requires a more 
complicated object, $\langle 0| \bar s(y)G_{\mu\nu}(x) b(0)|B_s(p)\rangle$, i.e., a generic 3DA with non-aligned coordinates. 

We perform the analysis using a field theory with scalar quarks/gluons which is technically very simple and allows one to 
focus on the conceptual issues; the generalization of our analysis for QCD is straightforward. We calculate nonfactorizable 
corrections directly, keeping control over all approximations. 
We adopt the counting scheme in which the parameter $\Lambda_{\rm QCD}m_b/m_c^2$ is kept of order unity, and show that the full 
3DA is necessary in order to resum properly the $(\Lambda_{\rm QCD}m_b/m_c^2)^n$ corrections: 
the dominant contribution to the $B$-decay amplitude are generated not only by the light-cone terms $y^2=0$ and $x^2=0$, 
but also by terms of order $\sim (xy)^n$. Therefore, the dominant contributions to the $B$-decay amplitude come from the 
configurations when both $x$ and $y$ lie on the light cone, 
but on the different axes: if $x$ is aligned along the $(+)$-axis, then $y$ is aligned along the $(-)$-axis. 

Expressing the $B$-decay amplitude via the standard 3DA with the aligned arguments, 
one can resum only a part of the $(\Lambda_{\rm QCD}m_b/m_c^2)^n$ corrections, 
whereas another source of the corrections of the same order 
remains unaccounted.

\section{Nonfactorizable corrections in a field theory with scalar particles}
\label{Sec_model}
In order to exemplify the details of the calculation, we consider nonfactorizable effects for the case of 
spinless particles. 
We shall use the standard QCD notations for spinor fields and assume that $m_b\gg  m_c \gg m_s\sim \Lambda_{\rm QCD}$, 
but the parameter $\Lambda_{\rm QCD}m_b/m_c^2$ is of order unity. 

\begin{center}
\begin{figure}[!hb]
\mbox{\epsfig{file=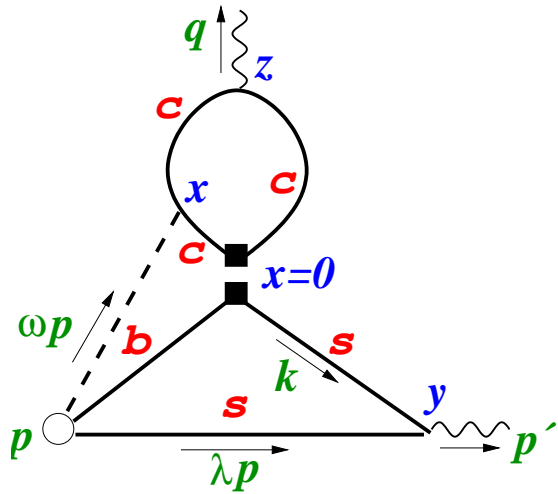,height=6.5cm}}
\caption{\label{Fig:1}
One of the diagrams describing the nonfactorizable gluon exchange. 
Dashed line corresponds to gluon; $q$ and $\kappa=-\omega p$ are the 
momenta outgoing from the charm-quark loop; the momentum $q'=q+\kappa=q-\omega p$ is emitted from the $b\to s$ vertex. 
Another diagram, equal to the one shown in the figure, corresponds to the 
Gluon attached to the right $c$-quark line in the upper loop.} 
\end{figure}
\end{center}

\noindent
We study the amplitude 
\begin{eqnarray}
\label{Apqoriginal}
A(p,q)=i\,\int dz e^{i q z}\langle 0|T\{c^\dagger(z)c(z),s^\dagger(0)s(0)\}|B_s(p\rangle, 
\end{eqnarray}
which involves weak interactions. We want to study nonfactorizable corrections due to a soft-gluon exchange 
between the charm-quark loop and the $B$-meson loop. To lowest order, the corresponding amplitude is given by 
the diagram of Fig.~\ref{Fig:1}:  
\begin{eqnarray}
\label{Apqscalar}
A(p,q)=i\,\int dz e^{iq z}\langle 0|T\{c^\dagger(z)c(z),
\,i\int dy' \,L_{\rm weak}(y'),
\,i\int dx\, L_{Gcc}(x), 
\,s^\dagger(0)s(0)\}|B_s(p\rangle, 
\end{eqnarray}
where the effective Lagrangian that mimics weak four-quark interaction is taken in the form 
\begin{eqnarray}
L_{\rm weak}=\frac{G_F}{\sqrt2}\, s^\dagger b\,c^\dagger c,
\end{eqnarray}
and the scalar gluon field $G(x)$ couples to the scalar $c$-quarks via the interaction  
\begin{eqnarray}
L_{\rm Gcc}=G\,c^\dagger c,
\end{eqnarray}
i.e., $G$ involves the quark-gluon coupling. 

First, we consider the charm-quark loop with the emission of a soft scalar gluon. We use the gluon field in momentum representation, 
related to the gluon field in coordinate representation as 
\begin{eqnarray}
G(x)=\frac{1}{(2\pi)^4}\int d\kappa \,\tilde G(\kappa)\,e^{i\kappa x},\quad \tilde G(\kappa)=\int dx \,G(x)\,e^{-i\kappa x}. 
\end{eqnarray}
Then the effective operator describing the gluon emission from the charm quark loop may be written as  
\begin{eqnarray}
\label{t1}
{\cal O}(q)=\int d\kappa\, \tilde G(\kappa)\,\Gamma_{cc}(\kappa,q),  
\end{eqnarray}
where $\Gamma_{cc}(\kappa,q)$ stands for the contribution of two triangle diagram with the charm quark running in the loop. 
The momenta $\kappa$ and $q$ are outgoing from the charm-quark loop, whereas the momentum $q'=q+\kappa$ 
is emitted from the $b\to s$ vertex. $p'$ is the momentum of the outgoing $s^\dagger s$ current and $p$ is the momentum of the $B$-meson, 
$p=p'+q$.

In terms of the gluon field operator in coordinate space, we can rewrite (\ref{t1}) as 
\begin{eqnarray}
\label{t2}
{\cal O}(q)=\int d\kappa\, e^{-i\kappa x}dx\,G(x)\Gamma_{cc}(\kappa, q), 
\end{eqnarray}

By virtue of (\ref{t2}), the amplitude Eq.~(\ref{Apqscalar}) takes the form
\begin{eqnarray}
\label{A}
A(q,p)&=&\frac{1}{(2\pi)^8}\int \frac{dk}{m_s^2-k^2}
\int dy e^{-i(k-p')y}
\int dx e^{-i\kappa x}d\kappa\, 
\Gamma_{cc}(\kappa, q)\,\langle 0|\bar s(y)G(x) b(0)|B_s(p)\rangle. 
\end{eqnarray}
Here, we encounter the $B$-meson three-particle amplitude with three different (non-aligned) arguments, 
for which we may write down the following decomposition: 
\begin{eqnarray}
\label{3DAnew}
\langle 0|s^\dagger(y)G(x) b(0)|B_s(p)\rangle=
\int d\lambda e^{-i \lambda y p}
\int d\omega e^{-i \omega x p}\,
\left[\Phi(\lambda,\omega)+O\left(x^2,y^2,(x-y)^2\right)\right],  
\end{eqnarray}
where $\lambda$ and $\omega$ are dimensionless variables. Making use of the properties of Feynman diagrams, 
it may be shown that they should run from 0 to 1. However, if one of the constituents is heavy, it carries the 
major fraction of the meson momentum and as the result the function $\Phi(\lambda,\omega)$ is strongly peaked 
in the region  
\begin{eqnarray}
\label{peaking}
\lambda, \omega=O(\Lambda_{\rm QCD}/m_b).  
\end{eqnarray}
So, effectively one can run the $\omega$ and $\lambda$ integrals from 0 to $\infty$; the latter integration limits 
emerge in the DAs within heavy-quark effective theory \cite{hidr,japan}. We emphasize that 
for the results presented below only peaking of the DAs in the region (\ref{peaking}) is essential. 
Notice also that the function $\Phi(\lambda,\omega)$ in (\ref{3DAnew}) coincides with the same 
function that appears in the ``standard'' 
3-particle distribution amplitude with the aligned arguments, $x=u y$, discussed in \cite{japan}.

\subsection{Light-cone contribution}

First, let us calculate the contribution to $A(q,p)$ from the term given by $\Phi(\lambda,\omega)$ in the 3DA (\ref{3DAnew}), i.e. 
corresponding to $x^2=y^2=(x-y)^2=0$. After inserting (\ref{3DAnew}) into (\ref{A}) we can perform the $x-$ and $y-$integrals
\begin{eqnarray}
\label{xyint}
&&\int dx\to\delta(\kappa +\omega p), \qquad \nonumber\\
&&\int dy\to\delta(k+\lambda p-p'). 
\end{eqnarray}
The next step is easy: the $\delta$-functions above allow us to take integrals over $k$ and $\kappa$, and we find
\begin{eqnarray}
\label{Aqp}
A(q,p)=\int_0^\infty d\lambda \int_0^\infty d\omega\, \Phi(\lambda,\omega)
\Gamma_{cc}\left(-\omega p, q \right)\frac{1}{m_s^2-(\lambda p-p')^2}.  
\end{eqnarray} 
For the sum of two triangle diagrams with the charm quark running in the loop we may use the representation
\begin{eqnarray}
\label{tFeyn}
\Gamma_{cc}(\kappa, q)=\frac{1}{8\pi^2}
\int\limits_{0}^{1}du \int\limits_{0}^{1-u}dv \,
\frac{1}{m_c^2-2uv \kappa q -u(1-u)\kappa^2-v(1-v)q^2}. 
\end{eqnarray}
Now, we must take into account that the $\omega$-integral is peaked at $\omega\sim \Lambda_{\rm QCD}/m_b$ so the gluon is soft:  
$\kappa=- \omega p$ and $\kappa^2\sim O(\Lambda_{\rm QCD}^2)\ll m_c^2$. 
The momentum transferred in the weak-vertex is $q'=q+\kappa=q-\omega p$, such that 
\begin{eqnarray}
q'^2=(q-\omega p)^2=q^2-\omega (1-\omega )M_B^2-q^2\omega +p'^2 \omega =q^2-\omega(1-\omega)M_B^2.
\end{eqnarray}
By virtue of the $y$-integration in (\ref{xyint}), the $s$-quark propagator takes the form  
\begin{eqnarray}
m_s^2-(\lambda p-p')^2=m_s^2-\lambda q^2+(1-\lambda)(\lambda M_B^2-{p'}^2).
\end{eqnarray}
Therefore, in the bulk of the $\lambda$-integration the virtuality of the $s$-quark propagator is large, $O(M_B)$. 
Let us notice that the $q^2$-dependence of the $s$-quark propagator is very mild and can be neglected; the main 
$q^2$-dependence of the amplitude $A(q,p)$ comes from the charm-quark loop.

\subsection{Deviations from the light-cone}
We now turn to the calculation of the contributions to $A(q,p)$ generated by terms $\sim x^2,y^2,(x-y)^2$ in the 3DA (\ref{3DAnew}). 
The terms containing powers of 4-vectors $y$ and $x$ in the integral (\ref{A}) can be calculated by parts 
integration leading to additional factors 
under the integrals:  
\begin{eqnarray}
y_\alpha\to 
\frac{k_\alpha}{\Lambda_{\rm QCD}m_b},\qquad
x_\alpha\to \frac{\{q_\alpha,\kappa_\alpha\}}{m_c^2}.
\end{eqnarray}
Taking into account the results (\ref{xyint}), we find the following relative contributions of the terms 
containing different powers of the coordinate variables: 
\begin{eqnarray}
y^2 &\to& \frac{k^2}{\Lambda_{\rm QCD}^2 m_b^2}\sim \frac{1}{\Lambda_{\rm QCD}m_b},\nonumber\\
x^2&\to& \frac{q\kappa}{m_c^4}\sim \frac{\Lambda_{\rm QCD} m_b}{m_c^4},\nonumber\\
xy&\to& \frac{(p'-\lambda p )(q-\omega p)}{\Lambda_{\rm QCD}m_b m_c^2}\sim \frac{m_b}{\Lambda_{\rm QCD}m_c^2}.
\end{eqnarray}
Clearly, all terms containing powers of $x^2$ and/or $y^2$ in the 3DA lead to the suppressed contributions to $A(q,p)$ 
and may be neglected within the considered accuracy.  
However, the terms containing powers of $xy$ lead to the contributions containing powers of ${\Lambda_{\rm QCD}m_b}/{m_c^2}$, 
i.~e., of order unity within the adopted counting rules. The kinematics of the process is thus rather simple: 
the vectors $x$ and $y$ are directed along the light-cone 
[e.g., $x$ along the $(+)$ axis, and $y$ along the $(-)$ axis], but the 4-vector 
$x-y$ is obviously not directed along the light cone. Therefore, the full dependence of 3DA on the variable $(x-y)^2$ 
is needed in order to properly resum corrections of order $\left(\Lambda_{\rm QCD}m_b/m_c^2\right)^n$. 

\section{Conclusions}
We have revisited the calculation of nonfactorizable charm-loop effects in rare FCNC $B$-decays.
To put emphasis on the conceptual aspects and to make the discussion clearer, we have considered the case of all scalar particles, 
avoiding in this way conceptually unimportant technical details. Our conclusions are as follows: 

\begin{itemize}
\item[(i)]
The relevant object that arises in the calculation of the nonfactorizable corrections is the three-particle DA with 
non-aligned coordinates: 
\begin{eqnarray}
\label{res1}
\langle 0|s^\dagger(y)G(x)b(0)|B_s(p)\rangle=
\int d\lambda e^{-i \lambda y p}
\int d\omega e^{-i \omega x p}\left[\Phi(\omega,\lambda)+O\left(x^2,y^2,(x-y)^2\right)\right]. 
\end{eqnarray} 
The function $\Phi(\omega,\lambda)$ here is precisely the same function that parameterizes the standard 3DA 
with the aligned arguments, $x=uy$, discussed in \cite{japan}.  
At small $q^2\le m_c^2$, terms of order $\sim x^2,y^2$ yield the suppressed contributions to the nonfactorizable 
amplitude of $B$-decay compared to the contribution of the light-cone term in the three-particle DA: 
for terms $O(x^2)$ the suppression parameter is $\Lambda_{\rm QCD}^2/m_c^2$, and for terms $O(y^2)$ 
the suppression parameter is $\Lambda_{\rm QCD}/m_b$. 
However, terms $\sim (xy)^n$ in the 3DA yield the contributions of order $\left(\Lambda_{\rm QCD}m_b/m_c^2\right)^n$ 
in the $B$-meson amplitude, i.~e., to the unsuppressed contributions. These contributions have the same 
order as the difference between the local OPE \cite{voloshin} and the light-cone OPE \cite{hidr} and should be 
properly resummed. The kinematics of the process looks simple: the 4-vectors $x$ and $y$ are directed along the light-cone 
[e.g., $x$ along the $(+)$-axis, and $y$ along the $(-)$-axis], but the 4-vector 
$x-y$ is obviously not directed along the light cone; therefore, the full dependence of the 3DA (\ref{res1}) 
on the variable $(x-y)^2$ is needed in order to properly resum corrections of order $\left({\Lambda_{\rm QCD}m_b}/{m_c^2}\right)^n$. 

\item[(ii)]
Evidently, a consistent treatment of nonfactorizable charm-loop effects in FCNC $B$-decays in QCD also requires 
the consideration of generic $B$-meson three-particle distribution amplitudes with non-aligned coordinates,
\begin{eqnarray}
\label{3DA_QCD}
\langle 0|\bar s(y)G_{\mu\nu}(x)b(0)|B_s(p)\rangle.
\end{eqnarray}
The Wilson lines between the field operators, making this quantity gauge-invariant, are implied. 
Notice that in QCD the complications of the generic 3DA (\ref{3DA_QCD}) compared to the 3DA with the aligned 
arguments are two-fold: 
(i) first, similar to the case of scalar constituents considered above, in each Lorentz structure that parametrizes (\ref{3DA_QCD}) 
one has to take into account terms $\sim (xy)^n$ that yield the unsuppressed contributions to the $B$-decay 
amplitude; (ii) second, (\ref{3DA_QCD}) contains 
additional Lorentz structures $\sim (x-y)_\alpha$ and  $\sim (x+y)_\alpha$ compared to the 3DA with the 
aligned arguments (see, e.g., Eq.~(4.7) of \cite{hidr}), and, respectively, new distribution amplitudes.  

\item[(iii)]
It seems plausible to infer that when considering non-factorizable gluon corrections in meson-to-vacuum 
transition amplitudes of the type $\langle 0|T \{j_1(z) j_2(0)\}|B\rangle$ 
(or similar amplitudes with light meson in the final state), one encounters two distinct kinds of processes: 

I. The amplitude of the process involves only one quark loop, i.e., the external boson is emitted 
from the same quark loop that contains valence quarks of the initial and the final mesons. In this case, non-factorizable corrections 
are light-cone dominated, i.e. may be expressed via light-cone three-particle distribution amplitude of the initial or 
of the final meson. For instance, weak form factors of $B$-meson decays treated within the method of light-cone sum rules 
(see e.g. \cite{lcsr1,lcsr2}) belong to this kind of processes. 

II. The amplitude of the process involves two separate quark loops (one quark-loop involving valence quarks of the initial 
and the final mesons and another quark loop that emits the external boson). In this case, the soft gluon from the initial or the final meson vertex 
is absorbed by a quark in a different loop. In this case, the description 
of non-factorizable soft-gluon corrections requires the full three-particle DA with non-aligned coordinates of the type 
of (\ref{res1}). Non-factorizable corrections to FCNC decays due to $c$- or $u$-quark loops belong to this kind of processes. 

A more detailed investigation of the general properties of non-factorizable corrections seems worthwhile.  
\end{itemize}

\acknowledgments{
D.~M. is grateful to H.~Sazdjian for illuminating discussions of the formalism of Wilson lines, to Y.-M.~Wang for a comment concerning 
his paper \cite{hidr}, and to W.~Lucha for interest in this work. D.~M. was supported by the Austrian Science Fund 
(FWF) under project P29028. Section 2 was done with the support of Grant No. 16-12-10280 of the Russian Science Foundation (A. K.)}


\begin{thebibliography}{100}
\bibitem{neubert}
M.~Beneke, G.~Buchalla, M.~Neubert, and C.~T.~Sachrajda,
{\it Penguins with Charm and Quark-Hadron Duality}, 
Eur.~Phys.~J. {\bf C61}, 439 (2009).

\bibitem{voloshin}
M.~B.~Voloshin, 
{\it Large $O(m_c^{-2})$ nonperturbative correction to the inclusive rate of the decay $B\to X_s\gamma$},  
Phys.~Lett. {\bf B397}, 275 (1997). 

\bibitem{buchalla}
G.~Buchalla, G.~Isidori, and S.~J.~Rey, 
{\it Corrections of order $\Lambda_{\rm QCD}^2/m_c^2$ to inclusive rare $B$ decays}, 
Nucl.~Phys. {\bf B511}, 594 (1998). 

\bibitem{khod1997}
A.~Khodjamirian, R.~Ruckl, G.~Stoll, and D.~Wyler, 
{\it QCD estimate of the long distance effect in $B\to K^*\gamma$}, 
Phys.~Lett. {\bf B402}, 167 (1997). 

\bibitem{zwicky1}
P.~Ball and R.~Zwicky, 	
{\it Time-dependent CP Asymmetry in $B\to K^*\gamma$ as a (Quasi) Null Test of the Standard Model}, 
Phys.~Lett. {\bf B642}, 478 (2006). 


\bibitem{zwicky3}
J.~Lyon and R.~Zwicky, 
{\it Resonances gone topsy turvy - the charm of QCD or new physics in $b\to s l^+l^-$?}
arXiv:1406.0566. 

\bibitem{ligeti}
Z.~Ligeti, L.~Randall, and M.~B.~Wise, 
{\it Comment on nonperturbative effects in $\bar B\to X_s\gamma$}, 
Phys.~Lett.~{\bf B402} 178 (1997). 

\bibitem{paz}
M.~Benzke, S.~J.~Lee, M.~Neubert, G.~Paz, 
{\it Factorization at Subleading Power and Irreducible Uncertainties in $\bar B\to X_s\gamma$}, 
JHEP {\bf 1008}, 099 (2010). 

\bibitem{hidr}
A.~Khodjamirian, T.~Mannel, A.~Pivovarov, and Y.-M.~Wang, 
{\it Charm-loop effect in $B\to K^{(*)} l^+l^-$ and $B\to K^*\gamma$}, 
JHEP {\bf 1009}, 089 (2010). 

\bibitem{japan}
H.~Kawamura, J.~Kodaira, C.-F.~Qiao, and K.~Tanaka, 
{\it B-meson light cone distribution amplitudes in the heavy quark limit}, 
Phys.~Lett.~{\bf B523}, 111 (2001), Erratum: Phys.~Lett. {\bf B536}, 344 (2002).

\bibitem{lcsr1}
P.~Ball and R.~Zwicky, 
{\it Improved analysis of $B\to\pi e\nu$ from QCD sum rules on the light cone}, 
JHEP {\bf 10}, 019 (2001). 

\bibitem{lcsr2}
A.~Khodjamirian, T.~Mannel, N.~Offen, 
{\it B-meson distribution amplitude from the $B\to\pi$ form-factor}, 
Phys.~Lett.~{\bf B620}, 52 (2005). 

\end{thebibliography}
\end{document}